\documentclass[12pt,preprint]{aastex}
\newcommand       \be           {\begin{equation}}
\newcommand       \ee           {\end{equation}}

\newcommand       \K            {\,{\rm K}}
\newcommand       \cm           {\,{\rm cm}}
\newcommand       \s            {\,{\rm s}}
\newcommand       \g            {\,{\rm g}}

\newcommand       \kms          {\,{\rm km \, s}^{-1}}

\newcommand       \yr           {\,{\rm yr}}

\newcommand       \gtsim        {\gtrsim}
\newcommand       \ltsim        {\lesssim}

\newcommand       \Tgas         {T_{\rm gas}}
\newcommand       \bomega       {\mbox{\boldmath$\omega$\unboldmath}}
\newcommand       \bB           {\mathbf{B}}
\newcommand       \bmu          {\mbox{\boldmath$\mu$\unboldmath}}
\newcommand       \ahat         {\hat{a}}
\newcommand       \rr           {\Upsilon}

\shortauthors{Weingartner}
\shorttitle{Grain Disalignment}

\begin{document}

\title{On the Disalignment of Interstellar Grains}

\author{Joseph C. Weingartner}
\affil{Department of Physics and Astronomy, George Mason University,
MSN 3F3, 4400 University Drive, Fairfax, VA 22030, USA}

\begin{abstract}

Several mechanisms have been proposed to explain the alignment of grains
with the interstellar magnetic field, including paramagnetic 
dissipation, radiative torques, and supersonic gas-grain streaming.  
These must compete with disaligning processes, including randomly
directed torques arising from collisions with gas atoms.  I describe a 
novel disalignment mechanism for grains that have a time-varying electric
dipole moment and that drift across the magnetic field.  Depending on the
drift speed, this mechanism may yield a much shorter disalignment timescale 
than that associated with random gas atom impacts.  For suprathermally 
rotating grains, the new disaligning process may be more potent for 
carbonaceous dust than for silicate dust.  This could result in efficient 
alignment for silicate grains but poor alignment for carbonaceous grains.

\end{abstract}


\section{\label{sec:intro} Introduction}

The phenomenon of starlight polarization was discovered over fifty years
ago (Hall 1949; Hall \& Mikesell 1949; Hiltner 1949a, 1949b)
and was quickly attributed to dichroic extinction by a population of
non-spherical, aligned interstellar dust grains.  In the ensuing decades,
many detailed investigations of potential alignment mechanisms were conducted, 
but to date no successful theory has been fully elaborated.  
See Lazarian (2003) and Roberge (2004) for recent reviews of 
grain alignment.  

Davis \& Greenstein (1951) developed a promising model based on dissipation
resulting from the rotation of a paramagnetic grain through the static
interstellar magnetic field $\mathbf{B}$.    
Since Fe is abundant and heavily depleted 
in the interstellar medium (ISM), it is likely that at least some grains
are paramagnetic.  Davis \& Greenstein assumed that the grain rotation is 
excited by elastic
impacts with gas atoms.  In this case, the energy in rotation about any
grain axis is $\sim \frac{1}{2}k\Tgas$, where $k$ is Boltzmann's
constant and $\Tgas$ is the gas temperature; thus, this motion is called
``thermal rotation''.

In addition to exciting grain rotation, the random torques imparted by 
gas atom impacts disalign grains.  Suppose each impact delivers angular
momentum $mva$, where $m$ is the mass of the gas atom, $v$ is its typical
speed, and $a$ is the grain size (which we will take to be the radius of 
a sphere of equal volume).  After $N$ collisions, the angular momentum 
axis will have suffered an angular displacement 
$\Delta \theta \sim (mva/J) N^{1/2}$; $J$ is the grain's total 
angular momentum.  For thermal rotation, $mv^2/2 \sim J^2/2I$, where
$I \sim Ma^2$ is the grain's moment of inertia ($M$ is the grain's mass), 
implying
$\Delta \theta \sim (Nm/M)^{1/2}$.  For $\Delta \theta \sim 1 \,$rad,
$N \sim M/m$.  Thus, the timescale for disalignment, $\tau_{\rm dis, \, col}$, 
equals the time for the grain to collide with its own mass in gas.
If the gas has number density $n$ and the grain, with 
mass density $\rho$, does not drift relative to the gas, then
\be
\tau_{\rm dis, \, col} \sim 1 \times 10^5 \, \left( \frac{\rho}{3 \g \cm^{-3}}
\right) \left( \frac{a}{0.1 \micron} \right) \left( \frac{n}
{30 \cm^{-3}} \right)^{-1} \left( \frac{T_{\rm gas}}{100 \K} \right)^{-1/2}
\yr~~~.
\ee
The disalignment timescale is expected to be shorter than the Davis-Greenstein
alignment timescale, unless the grains have superparamagnetic inclusions
(Jones \& Spitzer 1967).  

Thus, other alignment mechanisms were sought.  Martin (1971) pointed out that 
a charged, spinning grain has a magnetic dipole moment $\bmu$.  
Dolginov \& Mytrophanov (1976) showed that the Barnett effect
(i.e., the tendency for a spinning paramagnetic solid to 
acquire a magnetization parallel or anti-parallel to its angular velocity
$\bomega$) can provide a much larger moment.  Specifically, 
the Barnett magnetic moment $\bmu_{\rm Bar} = \chi_0 \bomega V/\gamma_g$, where
$\chi_0$ is the static magnetic susceptibility, $\gamma_g$ is the
gyromagnetic ratio of the microscopic magnetic dipoles that are responsible
for the grain's paramagnetism, and $V$ is the grain volume.  The torque
$\mathbf{\Gamma}_{\mu} = \bmu \times \bB$ causes the grain to precess
about the interstellar magnetic field $\bB$.  The precession rate
is fast enough that, even if the actual grain alignment mechanism did not
involve $\bB$, the observed polarization would be either parallel or
perpendicular to $\bB$.

Purcell (1979) noted that the Barnett effect also underlies a mechanism 
for dissipating grain rotational energy.  For fixed angular momentum 
$\mathbf{J}$, there are many rotational states available to a rigid 
body, with different rotational energies.  If the grain is not executing
steady rotation about one of its principal axes, then $\bomega$ varies
periodically in grain body coordinates.  The magnetization lags
$\bomega$ and the resulting paramagnetic dissipation, which Purcell
called ``Barnett dissipation'', transfers rotational energy to the 
thermal reservoir provided by the grain vibrational modes.  This drives
the principal axis of greatest moment of inertia, $\ahat_1$, into 
alignment with $\mathbf{J}$, since this configuration has minimum 
rotational energy for given $\mathbf{J}$.  Note that this mechanism acts 
even in the absence of an external magnetic field.

Purcell (1975, 1979) also noted that grains are subject to systematic torques
fixed in grain body coordinates and that these torques can spin the grains
up to suprathermal rotational speeds.  The most important torque identified
by Purcell results from the formation of H$_2$ at special surface sites,
with subsequent ejection from the grain.  This discovery restored the appeal
of the Davis-Greenstein mechanism, since suprathermally rotating
grains are largely impervious to disalignment by random gas atom impacts.
However, the distribution of H$_2$ formation sites on the grain surface
is expected to change (e.g., due to accretion of material from the gas)
on a timescale short compared with the alignment time.  Consequently, 
the associated torque will sometimes be oriented antiparallel to 
$\mathbf{J}$, spinning the grain down to the thermal rotation regime. 
Spitzer \& McGlynn (1979) found that grains are effectively disaligned
during these ``crossover'' episodes.  

Lazarian \& Draine (1997; 1999a, b) reexamined the physics of 
crossovers, including the effects of grain thermal fluctuations.
Thermal fluctuations, in which energy is spontaneously transferred from
the vibrational modes in the solid to grain rotation (at constant
$\mathbf{J}$), are associated with the internal dissipation mechanisms
(which we assume to be dominated by Barnett dissipation).  

Before proceeding, we must first summarize some key features of 
the torque-free rotational motion of asymmetric rigid bodies; see
\S 2.5 of Weingartner \& Draine (2003, hereafter WD03) for a more detailed 
description.  Suppose the inertia tensor has eigenvalues
$I_1 \ge I_2 \ge I_3$, with principal axes $\ahat_1$, $\ahat_2$, and
$\ahat_3$.
It is convenient to express the rotational
energy in terms of the following dimensionless parameter:
\be
q \equiv \frac{2 I_1 E}{J^2}~~~.
\ee
Note that $1 \le q \le I_1/I_3$.
The grain rotational state is not completely specified by
($\mathbf{J}$, $q$).  For any ($\mathbf{J}$, $q$), the grain is in one of
two possible ``flip states''.  These flip states are defined with respect
to one of the principal axes; which one depends on the value of $q$.
The grain is in the positive (negative) flip state with respect to
$\ahat_i$ if the component of the angular velocity along that axis,
$\omega_i$, remains positive (negative) throughout the grain rotational
motion.  

Lazarian \& Draine (1999a) identified an important consequence of thermal
fluctuations, which they called ``thermal flipping''.  Suppose a grain
initially has $q < I_1/I_2$ and is in the positive flip state with respect
to $\ahat_1$.  A thermal fluctuation may add enough energy to the rotational
motion so that $q > I_1/I_2$ (in which case the grain will be in one of the
flip states with respect to $\ahat_3$).  If the evolution subsequently
returns $q$ to a value $< I_1/I_2$, then the grain may end up in either
the positive or negative flip state with respect to $\ahat_1$; in the
latter case a thermal flip has occurred.  When a grain flips, any vector
fixed in grain-body coordinates, like the H$_2$ formation torque,
reverses direction in space.  Thermal fluctuations become more pronounced
as the grain angular momentum $\mathbf{J}$ decreases.  Lazarian \& 
Draine (1999a, b) found that thermally rotating grains with 
$a \ltsim 1 \micron$ undergo such rapid flipping that the H$_2$ formation
torque may never have the opportunity to spin a grain up to 
suprathermal rotation, a situation that they called ``thermal trapping''.
Thus, it appears that Davis-Greenstein alignment with 
H$_2$ formation torque-induced suprathermal rotation is not viable.

Draine \& Weingartner (1996)
found that radiative torques (due to the absorption and scattering of 
starlight by an irregularly shaped grain) can also yield suprathermal 
rotation.  If the interstellar radiation field is modestly anisotropic,
then (1) the radiative torque does not reverse direction on each grain 
flip, perhaps overcoming the problem of thermal trapping, and 
(2) radiative torques can directly align grains, on a shorter timescale than
that of the Davis-Greenstein mechanism (Draine \& Weingartner 1997).  
Thus, the radiative torque alignment scenario 
can be summarized as follows.  The radiative torque, averaged over
grain rotation and precession, drives the grain to suprathermal rotation 
and aligns $\mathbf{J}$ with respect to $\bB$.  On a much faster timescale,
Barnett dissipation aligns $\mathbf{J}$ with $\ahat_1$.  

Suppose a grain has an electric dipole moment $\mathbf{p}$, in
addition to its magnetic dipole moment, and that it drifts with velocity 
$\mathbf{v}$ relative to the gas and magnetic field.  This grain will 
experience a torque
\be
{\mathbf{\Gamma}}_p = \mathbf{p} \times \left( \frac{\mathbf{v}}{c} \times 
\bB \right) ~~~,
\ee
where $c$ is the speed of light.  Consider a coordinate system $x$, $y$, 
$z$ in which $\bB = B \hat{z}$ and 
$\mathbf{v} = v_{\parallel} \hat{z} + v_{\perp} \hat{y}$. 
The field axis
$\hat{z}$ also serves as the polar axis for spherical coordinates, 
with polar angle $\theta$ and 
with the azimuthal angle $\phi$ measured relative to $\hat{x}$.  
Suppose in addition that $\ahat_1$, $\bmu$, and $\mathbf{p}$ lie parallel or 
anti-parallel to $\mathbf{J}$:  $\bmu = \mu_J \hat{J}$ and 
$\mathbf{p} = p_J \hat{J}$; both $\mu_J$ and $p_J$ can be positive or negative.
Then, adding the torques due to the
magnetic and electric dipoles yields
\be
\label{eq:Gamma}
\mathbf{\Gamma} = \mu_J B \hat{J} \times [ \hat{z} + \rr \, \hat{x} ]
\ee
with 
\be
\label{eq:r}
\rr \equiv \frac{p_J \, v_{\perp}}{\mu_J \, c}~~~.
\ee
Thus, the grain precesses not about $\hat{z}$, but about another axis 
tilted at polar angle $\delta = \tan^{-1} |\rr|$.  The azimuthal angle of this
axis is $\phi =0$ ($\pi$) when $\rr>0$ ($\rr<0$).  
Also, the precession rate is increased by a factor
$(1+\rr^2)^{1/2}$.  

In general, $\bmu$ and $\mathbf{p}$ are not parallel or anti-parallel to 
$\mathbf{J}$.  How does this affect the dynamics?  The torques should be 
averaged over grain rotation, since this occurs on a much faster timescale 
than any other relevant process.  Denoting such averages with a bar,
\be
\label{eq:Gamma_mu_avg}
\bar{\mathbf{\Gamma}}_{\mu} = \bar{\bmu} \times \bB = 
\left( \frac{\mu}{\omega} \right) \bar{\bomega} \times \bB 
= \left( \frac{\mu}{\omega} \right) 
\frac{q \mathbf{J}}{I_1} \times \bB
\ee
(WD03, \S 3.2).  
If the Barnett effect is responsible for the magnetic dipole moment, 
then $(\mu/\omega) = \chi_0 V/\gamma_g$.  Note that we have 
ignored the component of the magnetization that is out of phase
with $\bomega$; this is justified in \S 3.2 of WD03.
Note also that $\bar{\mathbf{\Gamma}}_{\mu}$ is a factor $q$ larger than its
value when $\mathbf{J} \parallel \ahat_1$ and it does not depend on the
flip state.

Similarly,
\be
\bar{\mathbf{\Gamma}}_p = \bar{\mathbf{p}} \times \left( \frac{\mathbf{v}}{c}
\times \bB \right)~~~.
\ee
From \S 3.4 of WD03,
\be
\label{eq:p_bar}
\bar{\mathbf{p}}(q, \pm) = \pm \left( \mathbf{p} \cdot \ahat_i \right)
f_i(q) \hat{J}
\ee
where
\be
\label{eq:f_i}
f_i(q) = 
\left[ \frac{|I_i -I_3 q|}{I_1 - I_3} \right]^{1/2} \frac{\pi}
{2 F(\pi/2, k_i^2)}~~~,
\ee
the $\pm$ refers to the flip state, $i=1$ (3) when $q < I_1/I_2$
($q > I_1/I_2$), 
$F$ is the elliptic integral of the first kind (see eq.~15 of WD03), and
\be
k_i^2 \equiv \frac{(I_2 - I_3) (q - 1)}{(I_1 - I_2) (1 - I_3 q /I_1)}
\ee
when $q < I_1/I_2$ and its inverse when $q > I_1/I_2$.
  
\begin{figure}
\epsscale{1.00}
\plotone{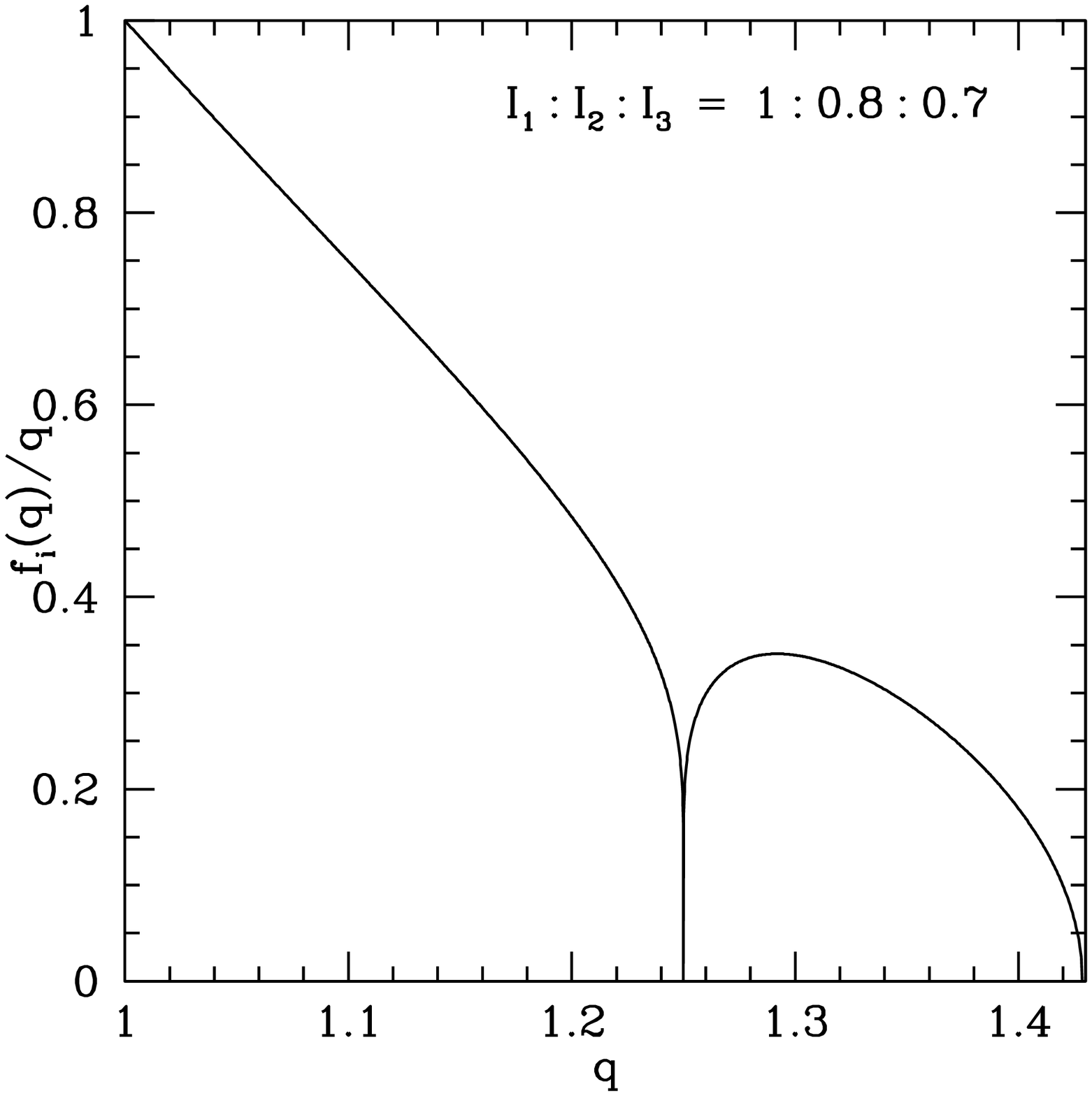}
\caption{
\label{fig:f_q}
$f_i(q)/q$, (see eq.~\ref{eq:f_i}),
with $i=1$ when $q < I_1/I_2$ and $i=3$ when $q > I_1/I_2$, for 
a grain with $I_1:I_2:I_3 = 1:0.8:0.7$.
        }
\end{figure}

Thus, if the grain rotational state remains fixed, then the dynamics is
identical to that for the case that $\mathbf{J}$, $\bmu$, and $\mathbf{p}$
all lie along $\ahat_1$ (eqs.~\ref{eq:Gamma} and \ref{eq:r}), with 
\be
\label{eq:mu_J}
\mu_J = \left(\frac{\mu}{\omega} \right) \frac{qJ}{I_1}
\ee
and
\be
\label{eq:p_J}
p_J = \pm \left( \mathbf{p} \cdot \ahat_i \right) f_i(q)~~~.
\ee
Figure \ref{fig:f_q} displays $f_i(q)/q$ for a grain 
with $I_1:I_2:I_3 = 1:0.8:0.7$.

The role of the electric dipole is insignificant (i.e., $\rr \ll 1$) when the 
grain rotates suprathermally and the drift speed perpendicular to the 
magnetic field is low.  Both of these conditions have traditionally been 
assumed, since grain alignment seemed to require suprathermal rotation and
gyrorotation and drag were thought to suppress drift across field lines.  

However, two recent studies call these assumptions into question.  
First, WD03 generalized the Draine \& Weingartner (1997) analysis of grain 
alignment by radiative torques.  Motivated by the fact that radiative 
torques drive suprathermal rotation, the earlier treatment adopted the 
assumption that $\mathbf{J} \parallel \ahat_1$.  However, radiative 
torques also naturally drive grains through crossovers; the importance of
thermal fluctuations and flips during these episodes invalidates this 
assumption.  Thus, WD03 dropped this requirement.  Surprisingly, they 
found that radiative torques can align $\mathbf{J}$ with respect to 
$\bB$ without driving the grain to suprathermal rotation.  For purposes
of computational speed, WD03 did not adopt a realistic spectrum for the
interstellar radiation field.  As a result, they could not estimate 
the timescale for alignment in the thermal rotation regime.  Thus, it 
is not yet clear whether the timescale for disalignment by random gas atom 
collisions is longer or shorter than the alignment time in this regime.

Second, Yan, Lazarian, \& Draine (2004) found that interstellar 
magnetohydrodynamic turbulence can accelerate grains to speeds as high
as $\sim 1 \kms$ across magnetic field lines in the cold neutral medium.
Thus, mildly supersonic motion is possible.  In this case, the angular
momentum imparted by collisions with gas atoms tends to lie perpendicular
to the drift velocity.  Yan et al.~(2004) suggested that the associated 
``mechanical alignment'' (Gold 1951, 1952) may be more important than 
previously thought.  Mildly supersonic drift also yields mildly 
suprathermal rotation.

\section{The Role of the Electric Dipole in Disalignment}

Suppose $p_J$ (eq.~\ref{eq:p_J}) is constant.  In this case, 
the only consequence of the electric dipole is to tilt the precession axis
relative to the magnetic field direction.  Because of gyrorotation, the 
precession axes of individual grains would likely be uniformly distributed 
in azimuth (about the field direction).    
Thus, the observed polarization is still either parallel or perpendicular
to the field, but is diluted.  Consider the simple case where the 
magnetic field lies in the plane of the sky and the grains are aligned such
that the precession 
angle (with respect to its instantaneous precession axis) is zero.  
For $\rr \ll 1$, the polarization is hardly affected.  As $\rr$ increases,
the dilution increases as well.  When $\rr=1$, the precession axis is tilted
$45\degr$ from the field direction.  Two grains that are $180\degr$ out
of phase in their gyrorotation produce orthogonal polarization vectors.
Thus, the polarization is completely suppressed when $\rr=1$.   As $\rr$ 
increases beyond 1, the dilution decreases, but does return to zero as
$\rr \rightarrow \infty$.  This polarization dilution is identical to that
which occurs when $\rr=0$ and the grains are aligned with a non-zero 
precession angle (with respect to the field direction).    

In the above paragraph, we assumed that $p_J$ is constant; in fact, 
it exhibits variations on a relatively 
short timescale.  Whenever the grain experiences a discrete charging 
event (e.g., the capture of an electron from the gas or the photoejection
of an electron), $\mathbf{p}$ 
will change.  It is not clear how the orientation of $\mathbf{p}$ in 
grain body coordinates is determined.  It may depend primarily on the
overall grain geometry.  Alternatively, it may be sensitive to localized
irregularities where charge can be concentrated.  In the former case,
the orientation of $\mathbf{p}$ is probably fairly constant.  In the 
latter case, it may change with each discrete charging event, and may 
even reverse direction with respect to $\ahat_1$ from time to time.  
Finally, $\mathbf{p}$ may harldy vary at all (in grain body coordinates).
This would be the case if it originates primarily in the random 
distribution of polar constituents rather than the asymmetric distribution 
of excess charge.

Even if $\mathbf{p}$ is fixed in grain body coordinates, $p_J$ will still
vary, if the rotational state (i.e., $q$ and/or the flip state) varies
(eq.~\ref{eq:p_J}).  The value of $q$ can be affected by both the 
action of external torques and internal relaxation processes, like Barnett
dissipation and fluctuations.  Of course, thermal flipping yields changes
in the flip state.  

The dynamical equation for $q$ is
\be
\label{eq:dq/dt}
\frac{dq}{dt} = \frac{1}{J^2} \left[ I_1 \left( \mathbf{J} \cdot
\frac{d\bomega}{dt} + \mathbf{\Gamma} \cdot \bomega \right) -
2 q \mathbf{J} \cdot \mathbf{\Gamma} \right]
\ee
(WD03, \S 2.4); internal relaxation is manifested in the 
$\mathbf{J} \cdot d\bomega/dt$ term.
The electric dipole does not affect $q$, since
$\overline{\mathbf{\Gamma}_p \cdot \bomega} = 0$ and
$\overline{\mathbf{J} \cdot \mathbf{\Gamma}_p} = 0$.
This would also be true for the magnetic torque if we ignore the lagging
component of magnetization, as in equation (\ref{eq:Gamma_mu_avg}).
The lagging component does yield a non-vanishing contribution, but it is
generally small compared with the Barnett dissipation rate for the grain
sizes considered here (\S 3.2 of WD03).
The contribution of other torques, like the radiative torque, are only 
significant on much longer timescales and can be ignored.
Thus, variations in $q$ are dominated by Barnett dissipation and
fluctuations, characterized by a timescale $\tau_{\rm Bar}$.
Of course, the thermal flipping timescale $\tau_{\rm tf}$ (i.e., the 
inverse of the probability per unit time that a flip occurs) must be
related to $\tau_{\rm Bar}$.

Variations in $\mu_J$ and $p_J$ (and hence $\rr$) cause the precession 
axis to vary, resulting in a potent mechanism for disaligning grains.
This process is most simply demonstrated if we assume that the magnitudes
of $\mu_J$ and $p_J$ are constant, but that the grain undergoes thermal
flipping, reversing the sign of $p_J$.  
Suppose that initially the grain is in the positive flip state with respect 
to $\ahat_1$
and $\mathbf{J}$ is precessing, with polar angle $\theta_1$, about its 
axis, tilted at angle $\delta$ relative to the magnetic field direction.  
Following a thermal
flip, the precession axis abruptly changes to the one associated with the
negative flip state.  The angle between these two axes is $2 \delta$.
The grain precesses about this new axis for some
time and then undergoes a thermal flip.  At that point, it resumes 
precessing about the original axis, but the polar angle now will not 
equal $\theta_1$.  After many flips, the grain's precession angle may
differ substantially from its initial value.

In order to estimate the disalignment timescale, it is convenient to 
write the equations of motion for $\hat{J}$ in the coordinate system 
($\theta$, $\phi$) with the magnetic field as the polar axis:
\be
\frac{d\phi}{dt} = \Omega_0 [ 1 - \rr \cot \theta \cos (\phi + 
\omega_{\rm gyro} \, t) ]
\ee
\be
\label{eq:dtheta_dt}
\frac{d\theta}{dt} = - \Omega_0 \rr \sin (\phi + \omega_{\rm gyro} \, t)~~~,
\ee
where $\Omega_0 \equiv - \mu_J B/J$ is the precession angular velocity 
in the absence of an electric dipole and $\omega_{\rm gyro}$ is the  
gyrofrequency.  In the limit $\tau_{\rm tf} \rightarrow 0$, the
terms proportional to $\rr$ average to zero and the grain simply precesses
about the magnetic field direction; $\theta$ remains fixed.
The timescale for gyrorotation is given by
\be
\omega_{\rm gyro}^{-1} \sim 2.4 \times 10^2 \left( \frac{\rho}{3 \g \cm^{-3}}
\right) \left( \frac{a}{0.1 \micron} \right)^2 \left( \frac{U}{0.3 \,
{\rm V}} \right)^{-1} \left( \frac{B}{5 \, \mu {\rm G}} \right)^{-1} \,
\yr~~~,
\ee
where $U$ is the grain potential.  For most cases of interest, this is 
much longer than the timescales for precession and flipping.

Suppose $\tau_{\rm tf}$ is much shorter than the precession timescale, 
$\Omega_{\Upsilon}^{-1} = |\Omega_0|^{-1} (1+\rr^2)^{-1/2}$.  
Consider the 
evolution over a time interval $dt_1 + dt_2$, during which the grain 
spends time $dt_1$ in the positive flip state and $dt_2$ in the negative
flip state.  From equation (\ref{eq:dtheta_dt}), the resulting deviation 
in polar angle 
$d\theta \sim |\Omega_0| \, \rr \, |dt_1 - dt_2| 
\sim |\Omega_0| \, \rr \, \tau_{\rm tf}$.
After $N$ double flips, the angular deviation $\Delta \theta \sim N^{1/2}
d\theta$.  For $\Delta \theta \sim 1 \,$rad, $N \sim d\theta^{-2}$.
Thus, the disalignment timescale 
\be
\label{eq:tau_dis_tf}
\tau_{\rm dis, \, tf} \sim N \tau_{\rm tf} \sim
\rr^{-2} |\Omega_0|^{-2} \tau_{\rm tf}^{-1}~~~~,~~{\rm if}~~~~~~\tau_{\rm tf} 
\ll |\Omega_0|^{-1} (1+\rr^2)^{-1/2}~~~.
\ee
In the opposite limit, 
where $\tau_{\rm tf} \gg \Omega_{\Upsilon}^{-1}$, the grain executes many full
precession cycles, plus one partial cycle, before flipping.  Since the 
deviation in $\theta$ over a full precession cycle equals zero, $d\theta$ 
is due to the difference in the final, partial precession cycles for the 
two flip states.  Thus,  
$d\theta \sim |\Omega_0| \, \rr \, \Omega_{\Upsilon}^{-1} = 
\rr (1+\rr^2)^{-1/2}$; 
this implies 
\be
\label{eq:tau_dis_tf2}
\tau_{\rm dis, \, tf} \sim (1+\rr^{-2}) 
\tau_{\rm tf}~~~~,~~{\rm if}~~~~~~\tau_{\rm tf} \gg
|\Omega_0|^{-1} (1+\rr^2)^{-1/2}~~~.
\ee
Note the opposite dependence of the disalignment timescale on the 
flipping timescale in the above two regimes; 
$\tau_{\rm dis, \, tf} \propto \tau_{\rm tf}$ ($\tau_{\rm tf}^{-1}$)
when flipping is slow (fast) compared with precession.

Of course, this scenario is simplified.  There are not sudden 
transitions between two values of $\rr$, with the same magnitude but 
opposite signs.  Rather, the grain experiences smaller variations in 
$\rr$ on a shorter timescale, $\tau_{\rm Bar}$.  Flipping is a cumulative
result of these variations.  We do not expect this complication to affect
the above order-of-magnitude
estimates for $\tau_{\rm dis, \, tf}$.  As long as the flip 
state remains the same, the sign of $d\theta$ will also remain the same
(in the limit that $\tau_{\rm tf} \ll \Omega_{\Upsilon}^{-1}$), 
even as $\rr$ varies.
Thus, it is still appropriate to consider the evolution between two 
flips to constitute a single step in the random walk.  The only modification
to the disalignment timescale is that $\rr$ should be replaced with 
$\langle \rr \rangle$, its thermal average over a distribution of 
rotational states (i.e., values of $q$) in a single flip state about
$\ahat_1$.  For a grain with temperature $T_d$, 
$I_1:I_2:I_3 = 1:0.8:0.7$, and 
$J^2/(2 I_1 k T_d) = 6$, $\langle \rr \rangle/\rr(q=1) \approx 0.7$; $\rr$ 
reaches its maximum value when $q=1$.  Thus, this modification is not very 
important.  (See \S 4.1 of WD03 for a derivation of the density of 
states as a function of $q$, which is needed to perform the thermal 
average.)  

Although a detailed model of Barnett relaxation and thermal
flipping has not yet been developed, Lazarian \& Draine (1999a) 
suggested that
\be
\label{eq:tau_tf}
\tau_{\rm tf}^{-1} \sim \tau_{\rm Bar}^{-1} \exp \left\{ - \frac{1}{2}
\left[ \left( \frac{J}{J_0} \right)^2 -1 \right] \right\}~~~,
\ee
where
\be
J_0 \equiv \left( \frac{I_1 I_2 k T_d}{I_1 - I_2} \right)^{1/2}~~~.
\ee
In the case of thermal rotation, $(J/J_0)^2 \sim T_{\rm gas}/T_d$;
thus $\tau_{\rm tf}$ is longer than
$\tau_{\rm Bar}$ by one to two orders of magnitude for thermally 
rotating grains in the cold neutral medium.

If discrete charging events substantially alter the orientation of 
$\mathbf{p}$ in grain body coordinates, then this process may also 
yield a form of flipping.  In this case, $\mathbf{p}$ flips in 
grain body coordinates rather than the grain orientation flipping 
with respect to $\mathbf{J}$.  The consequence for grain disorientation 
is the same.  If $\tau_{\rm dc}$ is the timescale on which discrete 
charging events occur and the fractional change in $p_J$ on each such 
event is $\sim Z^{-1}$ ($Z$ is the excess number of elementary charges
on the grain), then the associated ``flipping'' timescale
$\tau_{\rm f, \, dc} \sim Z^2 \tau_{\rm dc}$.  If 
$\tau_{\rm f, \, dc} < \tau_{\rm tf}$, then $\tau_{\rm f, \, dc}$ should
replace $\tau_{\rm tf}$ in the estimate of the disalignment time.

Thus, the disalignment timescale depends on (1) the drift speed $v_{\perp}$
across the magnetic field,  
(2)  the precession timescale $|\Omega_0|$,  
(3)  $\mu_J$ and $p_J$,
and (4)  $\tau_{\rm tf}$ (which is related to $\tau_{\rm Bar}$) or 
$\tau_{\rm f, \, dc}$.
In the next section, we will estimate the values of these quantities for
interstellar grains.

\section{Dipole Moments and Relaxation Timescales \label{sec:values}}

The magnetic dipole moment of a uniformly charged sphere with radius $R$,
total charge $q$, and rotational speed $\omega$ is $\mu_{\rm charge} = 
R^2 q \omega / 3c$.  Thus,
\be
\label{eq:mu_charge}
\mu_{\rm charge} \approx 1.1 \times 10^{-24} \, 
\left( \frac{U}{0.3 \, {\rm V}} \right) \left( \frac{a}{0.1 \micron} 
\right)^2 \left( \frac{\omega}{10^5 \s^{-1}} \right) \kappa_m
\, {\rm statcoulomb} \cm~~~,
\ee
where $U$ is the grain potential and $\kappa_m$ accounts for deviations from 
spherical shape.  Weingartner \& Draine (2001) estimated that 
$U \approx 0.3 \, {\rm V}$ for carbonaceous and silicate grains with 
$a \approx 0.1 \micron$ in the cold neutral medium (see their
fig.~10).  The thermal rotation rate for a sphere with radius $a$ is given by 
\be
\label{eq:omega_T}
\omega_T = \left( \frac{15 k \Tgas}{8 \pi \rho a^5} \right)^{1/2} = 
1.66 \times 10^5 \, \left( \frac{\rho}{3 \g \cm^{-3}} \right)^{-1/2} 
\left( \frac{\Tgas}{100 \K} \right)^{1/2} \left( \frac{a}{0.1 \micron}
\right)^{-5/2} \s^{-1}~~~,
\ee
where $\rho$ is the density of the grain material.  

The Barnett magnetic moment $\mu_{\rm Bar} = \chi_0 \omega V/\gamma_g$.
The susceptibility is given by Curie's Law (see, e.g., Morrish 1980):
\be
\label{eq:Curie}
\chi_0 = \frac{n \gamma_g^2 \hbar^2 J (J+1)}{3kT_d}~~~,
\ee
where $J$ is the angular momentum quantum number of the paramagnetic 
ion or nucleus, $n$ is its number density, and $T_d$ is the dust
temperature.

First, consider silicate grains with structural unit MgFeSiO$_4$.  The 
magnetization is dominated by the Fe ions.  Thus, we take $J=5/2$ and 
$\gamma_g = -1.76 \times 10^7 \s^{-1} \, {\rm G}^{-1}$, yielding
$\chi_0 \sim 5 \times 10^{-3} (T_d/15 \K)^{-1}$ (Draine 1996).
For silicates with less Fe incorporation,
$\chi_0$ would be lower. With these estimates,
\be
\label{eq:mu_Bar_sil}
|\mu_{\rm Bar}|({\rm sil}) \approx 
1.2 \times 10^{-19} \, \left( \frac{T_d}{15 \K}
\right)^{-1} \left( \frac{a}{0.1 \micron} \right)^3 \left( \frac{\omega}
{10^5 \s^{-1}} \right) \, {\rm statcoulomb} \cm~~~.
\ee

Next, consider carbonaceous grains with H incorporation at the $\approx 10\%$
level (i.e., the number density of H is $\approx 10^{22} \cm^{-3}$).  
The paramagnetism is dominated by the H nuclei, with
$\gamma_g = 2.675 \times 10^4 \s^{-1} \, {\rm G}^{-1}$ and $J=1/2$, implying
$\chi_0 \approx 9.6 \times 10^{-10} (T_d/15 \K)^{-1}$.  Thus, we estimate
\be
\label{eq:mu_Bar_carb}
\mu_{\rm Bar}({\rm carb}) 
\approx 1.5 \times 10^{-23} \, \left( \frac{T_d}{15 \K}
\right)^{-1} \left( \frac{a}{0.1 \micron} \right)^3 \left( \frac{\omega}
{10^5 \s^{-1}} \right) \, {\rm statcoulomb} \cm~~~.
\ee
If the carbonaceous grains are not hydrogenated to any significant extent,
then the nuclear
paramagnetism is contributed instead by $^{13}$C, which is present
at the $\sim 1\%$ level and has 
$\gamma_g = 6.73 \times 10^3 \s^{-1} \, {\rm G}^{-1}$ and $J=1/2$.
In this case, $\mu_{\rm Bar}$ would be reduced by a factor 
$n(^1{\rm H}) \gamma_g(^1{\rm H}) / n(^{13}{\rm C}) \gamma_g(^{13}{\rm C})
= 40$ compared with the above estimate.  
On the other hand, dangling bonds in 
hydrogenated amorphous carbon may yield electron paramagnetism with a spin
density $\sim 10^{19} \g^{-1}$ (Esquinazi \& H\"{o}hne 2005).
This is a factor $\approx 10^3$ less than the Fe density in the model 
silicate considered above.  Thus, the Barnett moment associated 
with these dangling bonds may exceed the estimate for hydrogenated 
carbon in equation (\ref{eq:mu_Bar_carb}) by an order of magnitude, 
depending on the value of $\gamma_g$.

If the electric dipole moment is due to an asymmetric distribution of excess
charge on the grain (characterized by an uknown parameter $\kappa_e$), then 
\be
\label{eq:p}
p = qa \kappa_e \approx 1.0 \times 10^{-15} \, \left( \frac{U}{0.3 \, {\rm V}}
\right) \left( \frac{a}{0.1 \micron} \right)^2 \left( \frac{\kappa_e}
{10^{-2}} \right) \, {\rm statcoulomb} \cm~~~.  
\ee
We have adopted a fiducial value of $\kappa_e \approx 10^{-2}$.  This
seems conservative, since the number of excess charges on a grain with 
$U = 0.3 \, {\rm V}$ and $a=0.1 \micron$ is only $\approx 20$.  In addition,
if the grain is composed of  $N$ randomly arranged polar constituents, 
each with dipole moment $p_0$, then we would expect an additional 
contribution to $p \sim N^{1/2} p_0$.  This may, in fact, dominate the 
contribution due to excess charge (see, e.g., eq.~11 in Draine \& Lazarian
1998).  

Observations of the wavelength dependence of starlight polarization reveal
that relatively large grains ($a \gtsim 0.1 \micron$) are aligned while
smaller grains ($a \ltsim 0.05 \micron$) are not (e.g., Kim \& Martin 1995).
For aligned grains, the Barnett magnetic moment dominates that due
to the spinning charge (compare eqs.~\ref{eq:mu_charge} and 
\ref{eq:mu_Bar_carb}).  For hydrogenated carbonaceous grains, the above
estimates yield
\be
\label{eq:compare}
\frac{pv_{\perp}}{\mu_{\rm Bar} c} \approx 220 \, 
\left( \frac{a}{0.1 \micron} \right)^{-1}  \left( \frac{\omega}{10^5 \s^{-1}} 
\right)^{-1} \left( \frac{T_d}{15 \K} \right) \left( \frac{U}{0.3 \, {\rm V}}
\right) \left( \frac{\kappa_e}{10^{-2}} \right)
\left( \frac{v_{\perp}}{1 \kms} \right)~~~.
\ee
Assuming thermal rotation (with $\Tgas = 100 \K$) 
and $a = 0.2 \micron$ (and that $T$, $U$, 
$\kappa_e$, and $v_{\perp}$ have the values in eq.~\ref{eq:compare}), 
$pv_{\perp}/\mu_{\rm Bar} c \sim 370$ for hydrogenated carbonaceous
grains and $\sim 0.046$ for silicate grains.  
Of course, these estimates are highly uncertain.

The precession rate
\be
\label{eq:prec}
|\Omega_0| = \frac{|\mu| B}{J} \approx 0.031 \,  \left( 
\frac{|\mu|}{10^{-23} \, {\rm statcoulomb \, cm}} \right) \left( \frac{B}
{5 \, \mu {\rm G}} \right) \left( \frac{\rho}{3 \g \cm^{-3}} \right)^{-1}
\left( \frac{a}{0.1 \micron} \right)^{-5} \left( \frac{\omega}
{10^5 \s^{-1}} \right)^{-1} \, \yr^{-1}~~~.
\ee
Since $\mu \propto \omega$, $\Omega_0$ is independent of $\omega$.

We adopt the order-of-magnitude estimate of the Barnett relaxation 
timescale from WD03 (their eq.~42).  With the additional assumptions
that $I_1 \approx 2 I_3$ and $I_3 \sim \frac{2}{5} \rho V a^2$, 
\be
\label{eq:tau_Bar1}
\tau_{\rm Bar}^{-1} \sim \frac{5 T_2}{\rho a^2} \frac{\chi_0}{\gamma_g^2}
\frac{\omega^2}{1 + \omega^2 T_1 T_2}~~~,
\ee
where $T_1$ and $T_2$ are the spin-lattice and spin-spin relaxation times,
respectively.  With equation (\ref{eq:Curie}),
this yields
\begin{eqnarray}
\label{eq:tau_Bar2}
\tau_{\rm Bar}^{-1} \sim 7 \times 10^3 \,
\left( \frac{n}{10^{22} \cm^{-3}} \right) \left( \frac{T_2}
{10^{-4} \s} \right) \left( \frac{a}{0.1 \micron} \right)^{-2} \left(
\frac{T_d}{15 \K} \right)^{-1}  \left( \frac{\rho}{3 \g \cm^{-3}} \right)^{-1}
\left[ \frac{J(J+1)}{3/4} \right]
\nonumber \\
\times \frac{100 \, \omega_5^2}{1 + 100 \,
\omega_5^2 \, (T_1 T_2 / 10^{-8} \s^2)}
\yr^{-1}~~~,
\end{eqnarray}
where $\omega_5 = \omega / 10^5 \s^{-1}$.  Note that 
$\tau_{\rm Bar} \propto T_2$ but does not depend on $\gamma_g$.  Thus,
the constituent that dominates the Barnett relaxation may be different 
than that which dominates $\mu_{\rm Bar}$.  
The fiducial values in equation (\ref{eq:tau_Bar2}) are all intended for
hydrogenated carbonaceous grains, for which Lazarian \& Draine (1999b)
estimated $T_1 \sim T_2 \sim 10^{-4} \s$.
For non-hydrogenated carbonaceous grains, $\tau_{\rm Bar}$ will be 
longer.  If silicate grains are hydrogenated, then equation 
(\ref{eq:tau_Bar2}) is applicable to them as well.  Otherwise, the 
$^{29}$Si nucleus (with $\gamma_g = -5.32 \times 10^3 \s^{-1} \, 
{\rm G}^{-1}$, $J=1/2$, and $n \approx 5 \times 10^{20} \cm^{-3}$)
will likely dominate Barnett relaxation;
this would increase $\tau_{\rm Bar}$ by a factor $\approx 20$.  For 
thermally rotating grains, the contribution of nuclear paramagnetism to
Barnett relaxation dominates that of electron paramagnetism, since
$T_2$ is orders of magnitude shorter for electron paramagnetism.

The discrete charging timescale can be simply approximated as the 
inverse of the rate at which electrons collide with the grain:
\be
\label{eq:tau_dc}
\tau_{\rm dc} \sim 1.5 \times 10^{-5} \, \left( \frac{a}{0.1 \micron}
\right)^{-2} \left( \frac{n_{\rm H}}{30 \cm^{-3}} \right)^{-1} \left(
\frac{x_e}{10^{-3}} \right)^{-1} \left( \frac{T_{\rm gas}}{100 \K} 
\right)^{-1/2} \left( \frac{1 + eU/kT_{\rm gas}}{36} \right)^{-1}
\, \yr~~~,
\ee
where $x_e$ is the electron fraction, $n_e/n_{\rm H}$, and the final 
term accounts for Coulomb focusing.    
We also estimate the timescale for the flipping of $\mathbf{p}$ in 
grain body coordinates (if discrete charging events can yield substantial
changes in $\mathbf{p}$)
as $\tau_{\rm f, \, dc} \sim Z^2 \tau_{\rm dc}$
with $Z \sim 20 (a/0.1 \micron)$ for $a \sim 0.1 \micron$.

\section{Consequences \label{sec:consequences}}

Consider hydrogenated carbonaceous and silicate
dust grains with $U \approx 0.3 \, {\rm V}$
in the cold neutral medium.  For the carbonaceous dust, we estimate 
$\langle \rr \rangle \sim 44 (a/0.1\micron)^{1.5} (\omega / \omega_T)^{-1}$ 
(i.e., 1/3 of the value 
estimated for $pv_{\perp}/\mu_{\rm Bar} c$ 
in \S \ref{sec:values}).  From equations
(\ref{eq:tau_tf}), (\ref{eq:tau_Bar2}), (\ref{eq:prec}), and 
(\ref{eq:tau_dc}), and assuming the fiducial values therein, 
\be
\label{eq:tau_Bar_carb}
\tau_{\rm Bar} \sim 5.2 \times 10^{-7} \left( \frac{a}{0.1\micron} 
\right)^7 \left( \frac{\omega}{\omega_T} \right)^{-2} \, 
\left[ 1 + 276 \left( \frac{a}{0.1\micron} \right)^{-5} \left( 
\frac{\omega}{\omega_T} \right)^2 \right] \, \yr~~~,
\ee
\be
\label{eq:tau_tf_carb}
\tau_{\rm tf} \sim \tau_{\rm Bar} \, \exp \left\{ \frac{1}{2} \left[ 6.7 
\left( \frac{\omega}{\omega_T} \right)^2 - 1 \right] \right\}~~~,
\ee
$|\Omega_0|^{-1} \sim 21 (a/0.1\micron)^2 \yr$ and 
$\tau_{\rm f, \, dc} \sim 6 \times 10^{-3} \yr$.
Similarly, for silicate dust
$\langle \rr \rangle \sim 0.0056 (a/0.1\micron)^{1.5} (\omega / \omega_T)^{-1}$
and $|\Omega_0|^{-1} \sim 0.0027 (a/0.1\micron)^2 \yr$.  The contribution 
of electron paramagnetism to $\tau_{\rm Bar}^{-1}$ can be ignored since
it is negligible for low to moderate $\omega/\omega_T$.  
For high $\omega/\omega_T$, 
where electron paramagnetism may be important for Barnett dissipation, 
the thermal flipping timescale is too long for the associated disalignment
to be significant.  Thus, we adopt equations (\ref{eq:tau_Bar_carb}) and
(\ref{eq:tau_tf_carb}) for silicate grains as well as for carbonaceous grains.

\begin{figure}
\epsscale{1.00}
\plotone{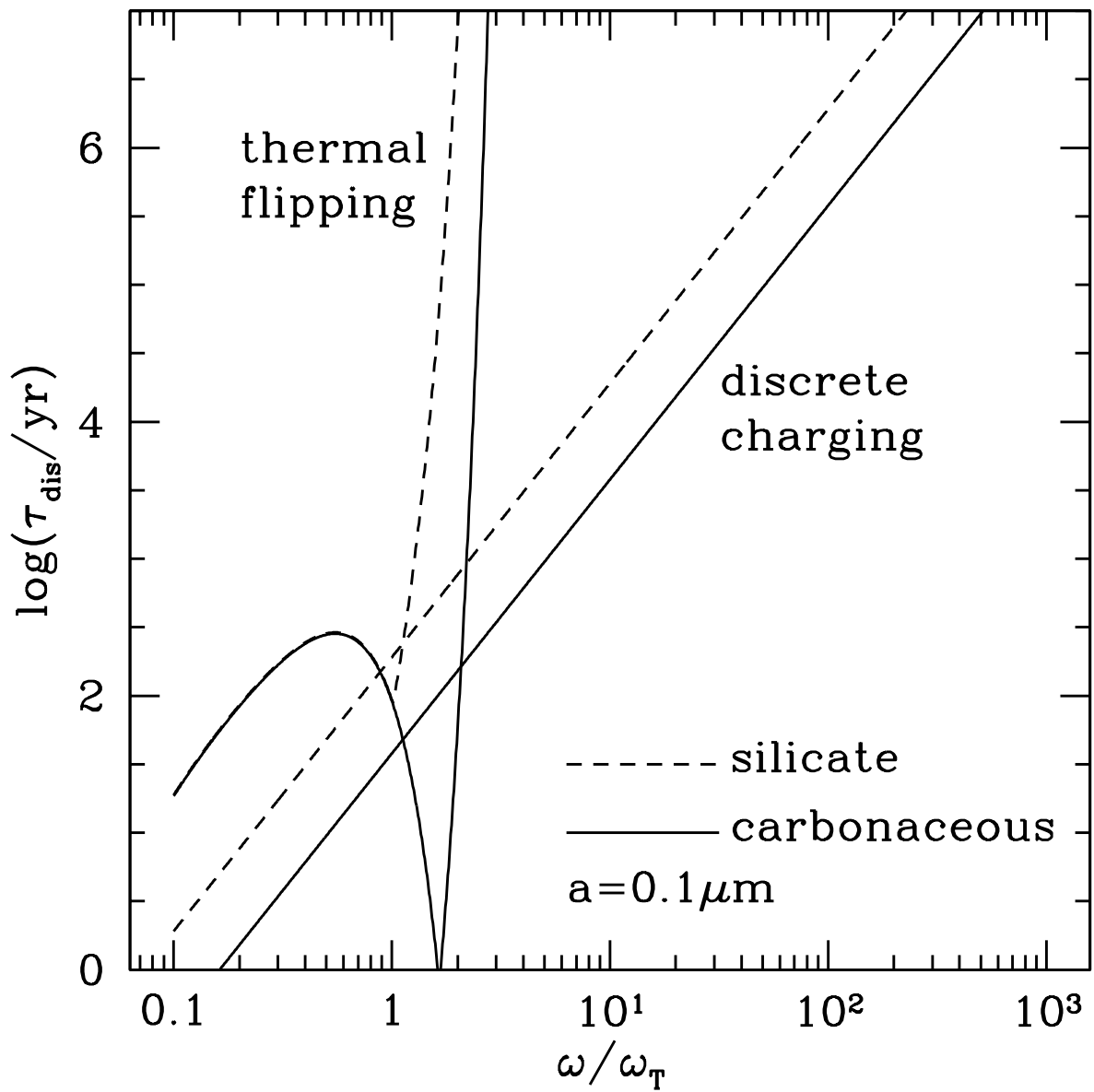}
\caption{
\label{fig:tau_dis_a0.1}
Disalignment timescales for changes in $\mathbf{p}$ associated with 
thermal flipping ($\tau_{\rm dis, \, tf}$) and discrete charging
($\tau_{\rm dis, \, dc}$), for carbonaceous and silicate grains with 
$a = 0.1 \micron$.  The suprathermality is indicated by the ratio of 
the angular speed $\omega$ to its thermal value, $\omega_T$.
        }
\end{figure}

\begin{figure}
\epsscale{1.00}
\plotone{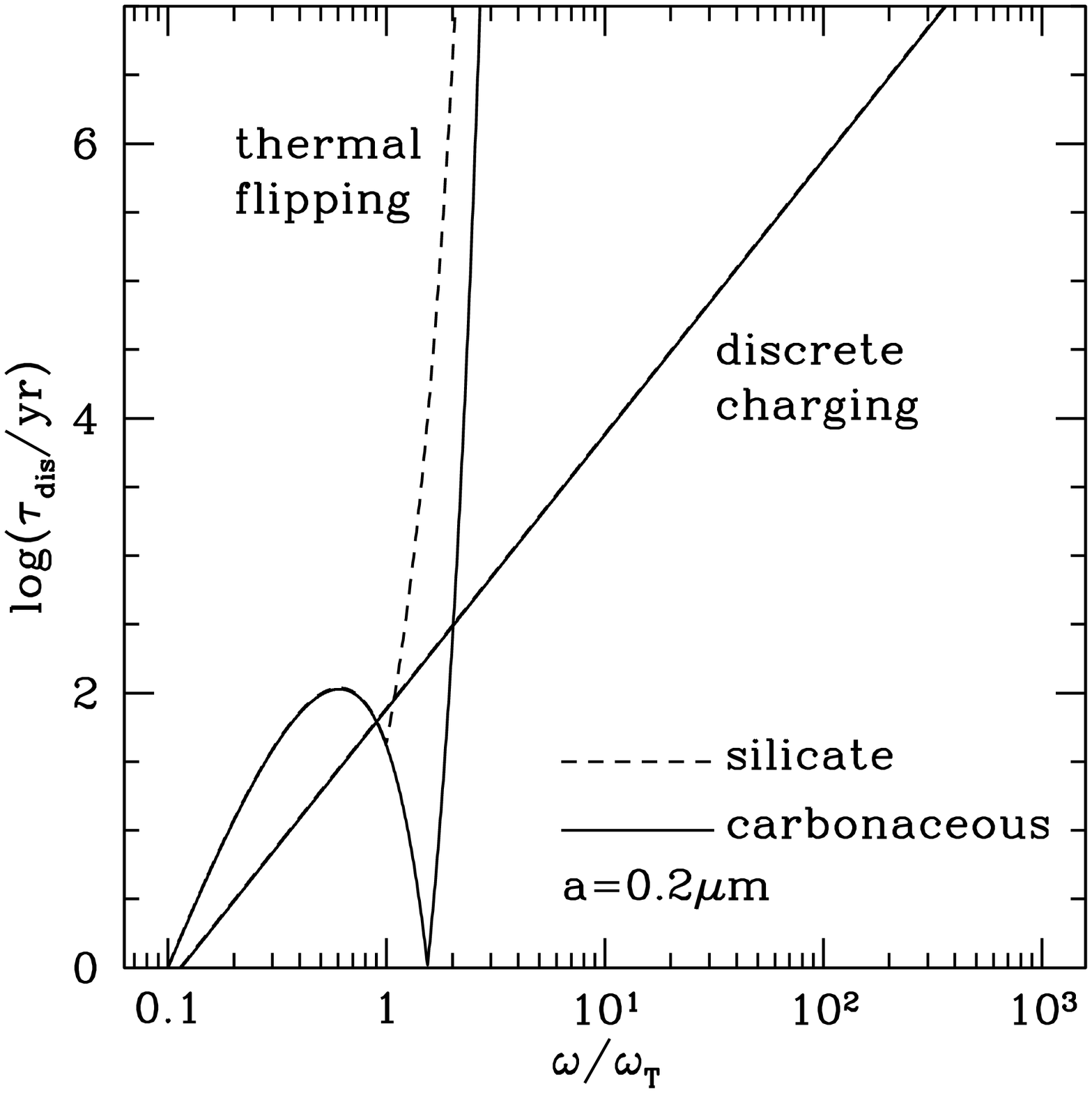}
\caption{
\label{fig:tau_dis_a0.2}
Same as fig \ref{fig:tau_dis_a0.1}, but for grains with $a=0.2 \micron$.
}
\end{figure}

We employ equation (\ref{eq:tau_dis_tf}) to estimate the timescale for
disalignment associated with thermal flipping, $\tau_{\rm dis, \, tf}$, 
when $\tau_{\rm tf} < |\Omega_0|^{-1} (1 + \rr^2)^{-1/2}$ and 
equation (\ref{eq:tau_dis_tf2}) when 
$\tau_{\rm tf} > |\Omega_0|^{-1} (1 + \rr^2)^{-1/2}$; similarly for the 
discrete charging disalignment timescale, $\tau_{\rm dis, \, dc}$.  
Figures \ref{fig:tau_dis_a0.1} and \ref{fig:tau_dis_a0.2} show 
$\tau_{\rm dis, \, tf}$ and $\tau_{\rm dis, \, dc}$ versus $\omega/\omega_T$,
for grains with $a=0.1 \micron$ and $0.2 \micron$, respectively.  
For the lowest values of $\omega$, $\tau_{\rm dis, \, tf}$ is identical for 
carbonaceous and silicate grains.  In this regime, 
$\tau_{\rm tf} < \Omega_0^{-1} (1 + \rr^2)^{-1/2}$ for both grain types. 
Although $\rr$
and $\Omega_0$ are very different for the two compositions, the 
disalignment timescale depends on the product $\rr \Omega_0$, in which the 
magnetic dipole moment cancels.  

For grains with $a=0.2 \micron$, the precession timescale always exceeds
$\tau_{\rm f, \, dc}$, for both carbonaceous and silicate grains.  Thus,
the discrete charging disalignment timescale is identical for the two 
grain types.  In contrast, $\tau_{\rm dis, \, dc}$ is a factor 
$\approx 5$ longer for silicate dust than for carbonaceous dust when 
$a=0.1 \micron$.  For silicate grains, $\tau_{\rm f, \, dc}$ exceeds 
the precession time; the opposite is true for carbonaceous grains.  
Thus, $\tau_{\rm dis, \, dc} = \rr_0^{-2} \tau_{\rm f, \, dc} (\omega/
\omega_T)^2$ for silicate grains and $\tau_{\rm dis, \, dc} = \rr_0^{-2}
\Omega_0^{-2} \tau_{\rm f, \, dc}^{-1} (\omega/\omega_T)^2$ for 
carbonaceous grains; $\rr_0 \equiv \rr(\omega = \omega_T)$.  

Suppose discrete charging events do not substantially change the 
orientation of $\mathbf{p}$ in grain body coordinates.  If the 
adopted parameter values are accurate, then thermally rotating grains of
both compositions are disaligned extremely rapidly.  The scenario of 
WD03, in which radiative torques align grains with $\omega \approx 
\omega_T$, would seem highly unlikely.  Alignment with suprathermal 
rotation may work, but disalignment associated with the electric dipole
would play a major role in the crossover dynamics.  Since 
$\tau_{\rm dis, \, tf} \propto (pv_{\perp})^{-2}$, 
errors in the assumed values 
of $p$ and $v_{\perp}$ could invalidate these conclusions.  Although the 
estimated value of $p$ seems conservative, a detailed model of the 
electric dipole is needed.  Also, confirmation that magnetohydrodynamic
turbulence can accelerate grains to $v_{\perp} \sim 1 \kms$ (Yan et al.~2004)
is needed.  Finally, a detailed model of thermal flipping is needed. 
If the steep increase in $\tau_{\rm tf}$ is shifted to somewhat lower
values of $\omega / \omega_T$, then alignment of thermally rotating 
grains may be tenable, even if $pv_{\perp}$ is large.  

If discrete charging events can lead to flips of $\mathbf{p}$ in grain
body coordinates, then even suprathermally rotating grains may be 
subject to significant disalignment.  The Davis-Greenstein alignment time
$\gtsim 10^6 \yr$ for $a \gtsim 0.1 \micron$ (see, e.g., eq.~71 in WD03).
Draine \& Weingartner (1997) found that radiative torques yield aligned grain 
states with $\omega / \omega_T \approx 100$ and alignment timescales 
$\gtsim 10^5 \yr$.
If the adopted estimate for $\tau_{\rm f, \, dc}$ is accurate, then 
Davis-Greenstein alignment with $\omega/\omega_{\rm T} \ltsim 100$ seems at 
best marginal, even if crossovers could be avoided.  Alignment by 
radiative torques fares better, but even in this case disalignment may 
play a significant role.  

Suppose the adopted value of the flipping timescale due to discrete 
charging, $\tau_{\rm f, \, dc}$, is too short.  When $a=0.1 \micron$, 
higher values of $\tau_{\rm f, \, dc}$ would yield higher values of 
the disalignment timescale $\tau_{\rm dis, \, dc}$ for silicate grains
(eq.~\ref{eq:tau_dis_tf}) and lower values of $\tau_{\rm f, \, dc}$ for 
carbonaceous grains (eq.~\ref{eq:tau_dis_tf2}).  If $\tau_{\rm f, \, dc}$
were sufficiently large that it exceeds the precession timescale for silicate 
grains with $a=0.2\micron$ (but not for carbonaceous grains), then 
disalignment 
would occur more rapidly for carbonaceous grains than for silicate grains
of this size.  If $\tau_{\rm f, \, dc}$ were larger than the adopted value 
by a factor of $\approx 40$, then $\tau_{\rm dis, \, dc} \gtsim 10^7 \yr$
for silicate grains, but $\tau_{\rm dis, \, dc} \ltsim 10^4 \yr$ for 
carbonaceous grains, when $a \approx 0.2 \micron$ and 
$\omega / \omega_{\rm T} \approx 10^2$.  
Thus, alignment may proceed more quickly than disalignment for silicate
grains, but vice versa for carbonaceous grains.  This could explain the 
observation that silicate grains are aligned while carbonaceous grains 
may not be (Whittet 2004).
Again, a detailed model of the electric dipole moment in interstellar
grains is needed to clarify this possibility.

With the above estimates, $\langle \rr \rangle \sim 1$ for carbonaceous
grains with $a \approx 0.2 \micron$ and $\omega/\omega_T \approx 100$.
Thus, even if these grains do undergo efficient alignment, the 
polarization may be significantly diluted.  

\section{Summary}

\noindent 1.  When a grain with an electric dipole moment drifts across the 
magnetic field $\bB$, the grain precesses about an axis different from 
$\bB$.

\noindent 2.  The importance of the electric dipole torque scales with the
parameter $\rr \equiv p_J \, v_{\perp} / \mu_J \, c$; 
$\rr$ is expected to be larger
for carbonaceous grains than for silicate grains.  

\noindent 3.  Because of gyrorotation, polarization still correlates with 
$\bB$ when $\rr > 0$, but the polarization is diluted, with maximal dilution 
when $\rr \approx 1$.  

\noindent 4.  Variation of the sign of $\rr$, due to thermal flipping or 
discrete charging, produces grain disalignment.  The timescale for this
disalignment process may be much shorter than that for disalignment by 
random collisions with gas atoms.

\noindent 5.  In the case of suprathermal rotation, the disalignment 
timescale may be shorter for carbonaceous grains than for silicate grains.

\noindent 6.  A detailed model of the electric dipole moment in 
interstellar grains is needed before firm estimates of the disalignment
timescale can be made.

\acknowledgements

Support for this work, part of the Spitzer Space Telescope Theoretical
Research Program, was provided by NASA through a contract issued by the
Jet Propulsion Laboratory, California Institute of Technology under a
contract with NASA.
I am grateful to Paula Ehlers for discussions that motivated this work, 
to David Barr for helpful discussions, and to Bruce Draine for helpful 
comments on a draft of the paper.

\end{document}